\documentclass[nohyper]{tufte-handout}

\title[Social Complexity Lab: Preprint\hfill Fundamental structures]{Fundamental structures in \\dynamic communication networks%
}

\author[Sune Lehmann]{Sune Lehmann%
\thanks{DTU Compute\\%
Technical University of Denmark\\%
DK-2800 Kgs Lyngby\\%
Denmark.\\%
Email: \textsf{sljo@dtu.dk}
}}

\date{July 2019} 

\usepackage{graphicx} 
  \setkeys{Gin}{width=\linewidth,totalheight=\textheight,keepaspectratio}
  \graphicspath{{graphics/}} 
\usepackage{amsmath}  
\usepackage{booktabs} 
\usepackage{units}    
\usepackage{multicol} 
\usepackage{lipsum}   
\usepackage{fancyvrb} 
  \fvset{fontsize=\normalsize}

\newcommand{\classtxt}[1]{\textit{\textbf{#1}}}
\usepackage{pdfpages}
\usepackage[draft]{hyperref}

\begin{document}

\includepdf{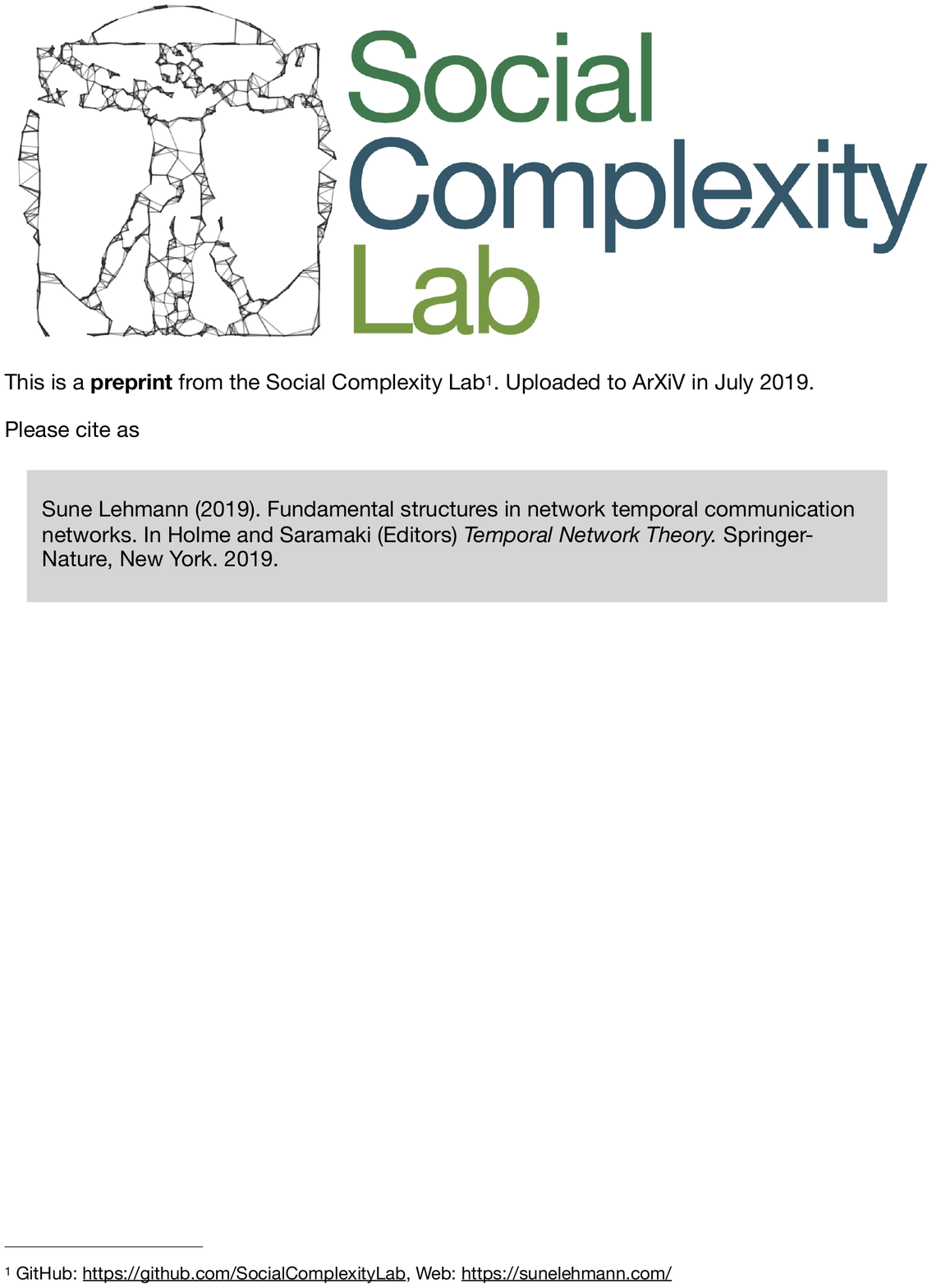}

\maketitle
\setcounter{page}{1} 
\begin{abstract}
\noindent
In this paper I introduce a framework for modeling temporal communication networks and dynamical processes unfolding on such networks. 
The framework originates from the realization that there is a meaningful division of temporal communication networks into six dynamic classes, where the class of a network is determined by its generating process. 
In particular, each class is characterized by a fundamental structure: a temporal-topological network motif, which corresponds to the network representation of communication events in that class of network. 
These fundamental structures constrain network configurations: only certain configurations are possible within a dynamic class. 
In this way the framework presented here highlights strong constraints on network structures, which simplify analyses and shape network flows. 
Therefore the fundamental structures hold the potential to impact how we model temporal networks overall. 
I argue below that networks within the same class can be meaningfully compared, and modeled using similar techniques, but that integrating statistics across networks belonging to separate classes is not meaningful in general.
This paper presents a framework for how to analyze networks in general, rather than a particular result of analyzing a particular dataset. 
I hope, however, that readers interested in modeling temporal networks will find the ideas and discussion useful in spite of the paper's more conceptual nature.
\end{abstract}

\section{Introduction}
Temporal networks provide an important methodology for modeling a range of dynamical systems\cite{holme2012temporal, holme2015modern}.
A central category of temporal networks is \emph{communication networks}, which -- in this context -- I define to be networks that facilitate or represent communication between human beings.
Frequently analyzed examples of communication networks are networks of face-to-face contacts between individuals, phone calls and text messages, online social networks such as Facebook or Twitter, and networks of email messages.
But communication networks could also represent other types of human communication, such as broadcast networks (e.g.~television  or newspapers) or communication via letters or books.
While the framework discussed here is presented in the context of human communication networks, in many cases the validity of the framework extends beyond networks of human communication to describe networks of machine-machine communication, biological signaling, etc.

\section{Network structure of communication events}
The main realization underlying the ideas presented here is that each human act of communication is shaped by the medium in which it takes place. 
As modern communication tools have developed, the richness of the ways human beings can communicate with one another has grown. 
What is perhaps less recognized in the field of network theory is that each new medium for communication sets its own particular constraints for the network structure of communication events within that medium.

In the field of communication studies, a key question is to understand how the technological evolution impacts human communication. 
Therefore, within that field, the many possible types of human communication -- old and new -- have been boiled down to six fundamental prototypical communicative practices\cite{jensen2011internet} shown in Table\,\ref{tab:matrix}.
In their formulation within the field of communication these practices are not connected to the underlying communication networks (or their dynamics); rather, the prototypical practices are simply used as a way to categorize real-world communication and understand their impact on, e.g.~communication practices.
\begin{table}
\begin{tabular}{p{1.9cm}|p{3.70cm}|p{3.70cm}}
   & \textbf{Synchronous} & \textbf{Asynchronous}\\\hline
\textbf{One-to-one} & Phone call, voice chat & Text message, letter \\\hline
\textbf{One-to-many} & Broadcast Radio and TV & Book, Newspaper, Webpage \\\hline
\textbf{Many-to-many} & Face-to-face, Online chatroom & Online social network, Wiki\\
\end{tabular}
\caption{Six prototypical communicative practices and real-world examples of each practice.  \label{tab:matrix}}
\end{table}

\subsection{\classtxt{Synchronous} vs.~\classtxt{Asynchronous}}
In the vertical split, Table\,\ref{tab:matrix} makes a distinction between \classtxt{synchronous} and \classtxt{asynchronous} communication. 
In the case of \classtxt{synchronous} communication, both parties are active and engaged. E.g.~during a phone call. 
Conversely, in the case of \classtxt{asynchronous} communication, a message is initiated at some time by the sender and then received at some later time by the recipient(s). 
For example, in the case of \classtxt{one-to-one} communication, the recipient reading a text message or a letter.

\subsection{\classtxt{One-to-one}, \classtxt{One-to-many}, \classtxt{Many-to-many}}
Along the horizontal splits, each row in Table\,\ref{tab:matrix} refers to the configuration of participants in a given communication act and the nature of their interaction. 
This division of communicative behaviors into \classtxt{one-to-one}, \classtxt{one-to-many}, and \classtxt{many-to-many} is quite natural and recognized beyond communication theory; similar distinctions are used, for example, in the analysis of computer networks\cite{carlberg1997building}, when negotiating contracts\cite{rahwan2002intelligent}, within marketing\cite{gummesson2004one}, or as design patterns/data models in database design\cite{jewettt2011database}.

\subsection{Connecting to network theory}
Bringing this framework, which was developed to organize different types of communication, into the realm of temporal network theory, I propose that we think of each prototypical type of communication as defining a \textit{\textbf{dynamic class}} of network and that any real-world communication network can be modeled as belonging to one of these six classes.

\begin{marginfigure}
	\includegraphics{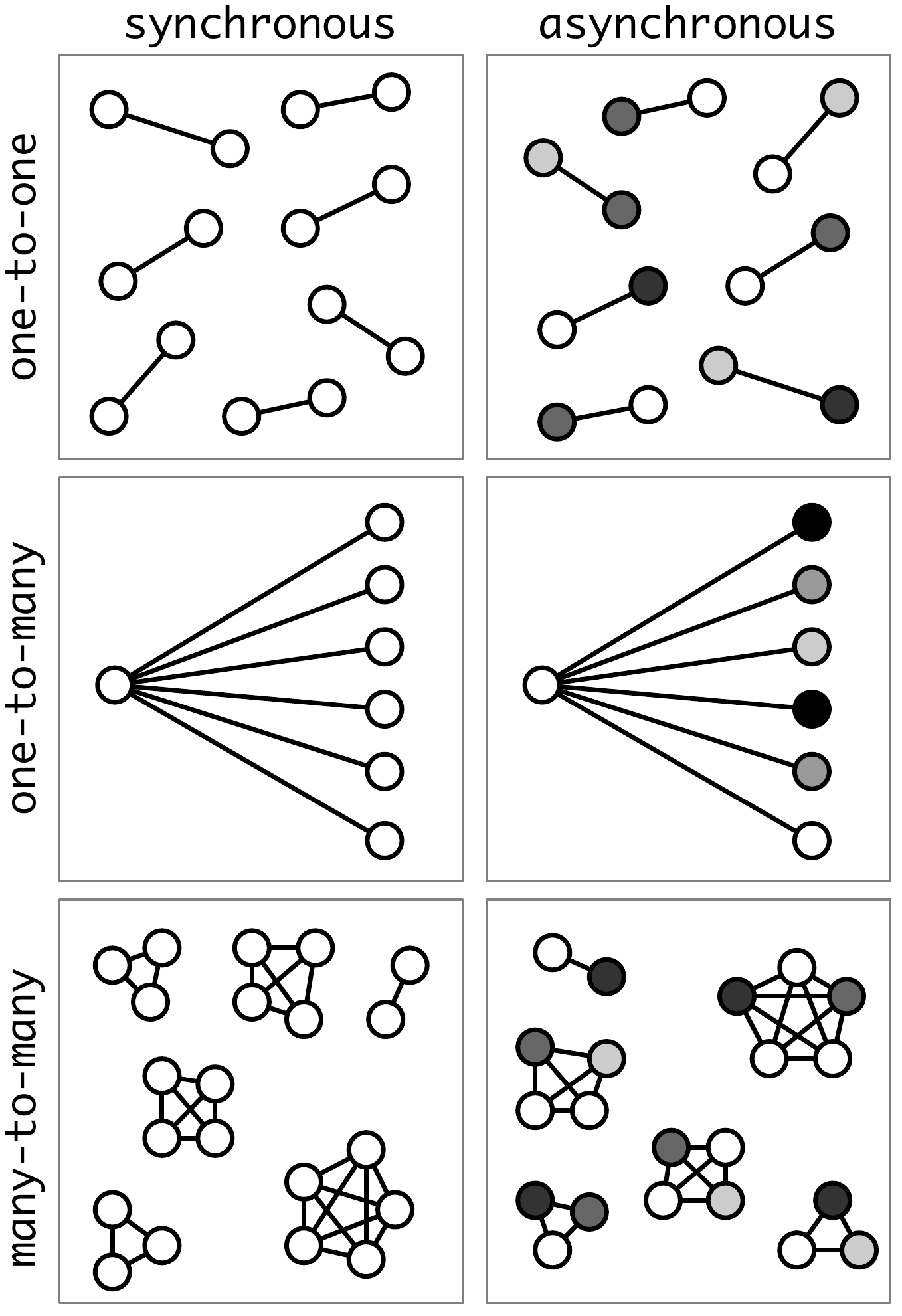}
	\caption{Network cross sections of all six classes fundamental structures (cf.~Table\,\ref{tab:matrix}). In the \classtxt{asynchronous} classes, I have colored nodes to illustrate the temporal dimension (older nodes are represented using darker colors). \label{fig:fundamental}}
\end{marginfigure}

The key concept which distinguishes the six classes is their \textit{\textbf{fundamental structures}}. 
We arrive at the fundamental structures by first noticing that each row in Table\,\ref{tab:matrix}, corresponds to an archetypal network structure: \classtxt{one-to-one} interactions correspond to \textit{dyads}, \classtxt{one-to-many} interactions can be represented as \textit{star graphs} (or \textit{trees}), and the \classtxt{many-to-many} interactions match the network structure of \textit{cliques}.
When also incorporating the temporal aspect (\classtxt{synchronous}/\-\classtxt{asynchronous}), we arrive at the network representations of the six prototypical communicative practices, the fundamental structures, see Figure\,\ref{fig:fundamental} for an illustration.

The fundamental structures are temporal-topological network patterns, with each pattern corresponding to a communication event (a phone call, a meeting, a text message) in that network.
Since each class is characterized by its fundamental structure, we name the each class according to their fundamental structure: \classtxt{synchronous, one-to-one}, and so on. 
\begin{figure}
\centering
\includegraphics[width=0.9\textwidth]{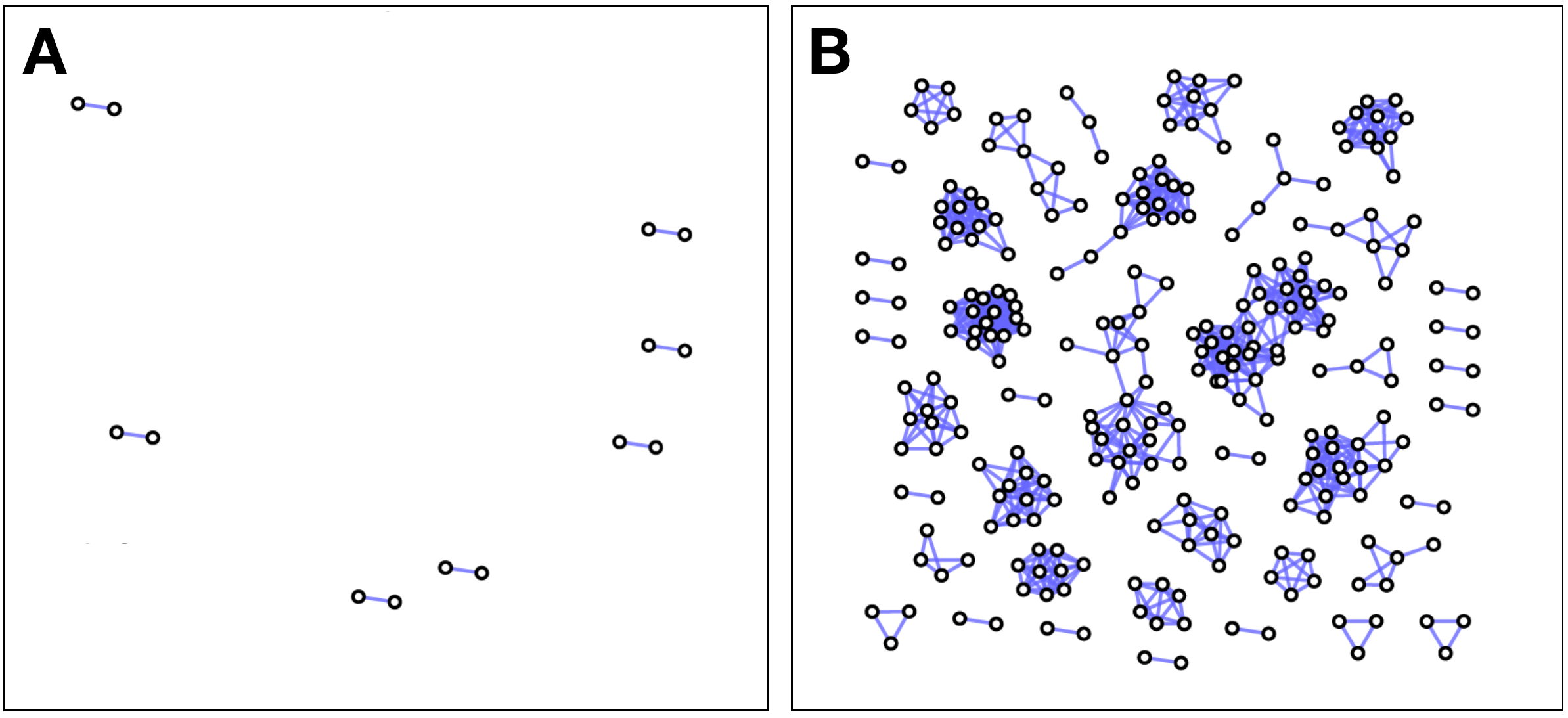}
\caption{Cross-sections of fundamental structures are revealed in brief snapshots of networks from the SensibleDTU project.
\textbf{A}.~A one minute time-slice from the phone call network at peak activity; the network is entirely composed of dyads.
\textbf{B}.~Social interactions over 5 minutes in the face-to-face contact network. Here the network is disconnected and well-approximated by non-overlapping cliques. \label{fig:examples}}
\end{figure}

Let me provide some examples to give a sense of what I mean.
In the \classtxt{synchronous}, \classtxt{one-to-one} class (e.g.~the phone call network), fundamental structures are individual dyads, with some duration given by its start and end time.
In \classtxt{synchronous}, \classtxt{one-to-many} networks (e.g.~a live-stream), the duration of the communication event is set by the node which is broadcasting, whereas receiving nodes may participate for only part of the communication event's duration.
Finally, in the \classtxt{synchronous}, \classtxt{many-to-many} class, where the fundamental structure is a sequence of cliques, the start of the communication event is set by the first participant(s) connecting -- and the end occurs when the last participants(s) stop communicating. An example of this class is face-to-face networks, where a fundamental structure could represent a group of friends meeting for dinner at a restaurant.

\newthought{In all of the} \classtxt{synchronous} classes, infinitesimally thin temporal slices of the communication event reveal the network pattern characteristic of that class. That is, a \textit{dyad}, \textit{tree}, or \textit{clique} for the \classtxt{one-to-one}, \classtxt{one-to-many}, and \classtxt{many-to-many} class respectively.
I illustrate this point in Figure\,\ref{fig:examples}, where I show brief snapshots (thin temporal slices) of the network of phone-calls (\classtxt{synchronous}, \classtxt{one-to-one}) which consists of disconnected dyads, each dyad corresponding to an ongoing conversation (see Figure\,\ref{fig:examples}A), whereas a slice of face-to-face meetings is well approximated as disconnected cliques (see Fig.\,\ref{fig:examples}B). 
Getting a bit ahead of myself, I note that, already at this point, it is clear from inspection that from the point of view of a dynamical process, the possible network flows in the two networks shown in Figure\,\ref{fig:examples} are going to be very different.
\marginnote[2mm]{Summarizing the discussion, I introduce two key concepts here. (1) \textbf{Dynamic Classes}: Each dynamic class is the set of networks characterized by a certain type of fundamental structure. There are six dynamic classes.
(2) \textbf{Fundamental structures}: A fundamental structure is the topological-temporal network representation the archetypical communication pattern within a class of network. 
Each communication event corresponds to an instance of the fundamental structure characterizing that network. 

A useful way to think about real-world communication networks is as a sequences of instances of fundamental structure from a single class.
In this sense we can think of each of the fundamental structures as \textit{generating} a \textit{class} of networks. }

Next, let us consider examples of networks from the \classtxt{asynchronous} classes. 
Here \classtxt{one-to-one} communication events still involves dyads, and an event starts, when a person sends a message.  
The communication ends when the recipient receives the message at some later time.
In the {asynchronous}, \classtxt{one-to-many} class, a communication event starts when some communication is initiated (and that node becomes \textit{active}): a book is published, a web-page is launched, etc. 
Now, recipients can engage with that active node at any point until the sender-node is no longer active/available -- and thus ending that fundamental structure.
Finally the \classtxt{asynchronous}, \classtxt{many-to-many} class. 
Here, again a node becomes active (starting the communication event), and other nodes can engage with the active node \textit{as well as all other nodes in that conversation}. 
The fundamental structure ends when original post ceases to be available (although activity may end sooner than that).
Examples of networks from the \classtxt{asynchronous}, \classtxt{many-to-many} class is a post on a message board, or a post on Facebook\footnote[2mm]{It is not guaranteed that all posts become a full discussion between all readers. And if no-one comments, such posts could display a \classtxt{one-to-many} structure. I discuss this below. }.
In the case of the  \classtxt{asynchronous} classes, infinitesimally thin time-slices of the fundamental structures are empty, only revealing the active nodes, since the interactions themselves typically are instantaneous and do not have a duration.

\newthought{The concept of classes} will prove to be important because while two networks originating from different classes aggregated over time may have similar topological properties, a difference in network class may have profound impact on the network dynamics and for processes unfolding on the network.  
Stated differently: \emph{When we consider the networks on much shorter time-scales than those typically considered in the literature, networks from the six classes are radically different}, cf.~Figure\,\ref{fig:fundamental}. 

I cover this point in detail below, arguing that there are a number of advantages associated with thinking about (and explicitly modeling) temporal networks as sequences of fundamental structures. 
Further, I argue below that when we compare analyses of various real-world networks, we should only expect similar behavior when we compare networks within the same dynamic class, and that we should aggregate statistics within each class of network separately.

\subsection{The case of \classtxt{many-to-many}, \classtxt{synchronous} networks}
Before we move on, let me start by showing how the fundamental structures can lead to clean, simple descriptions of temporal communication networks. 
A few years ago\cite{sekara2016fundamental} -- without realizing the connection to a larger framework -- my group analyzed a network from the class of \classtxt{synchronous}, \classtxt{many-to-many} interactions, arising from person-to-person interactions in a large, densely connected network.
\begin{figure}
\centering
\includegraphics[width=0.9\textwidth]{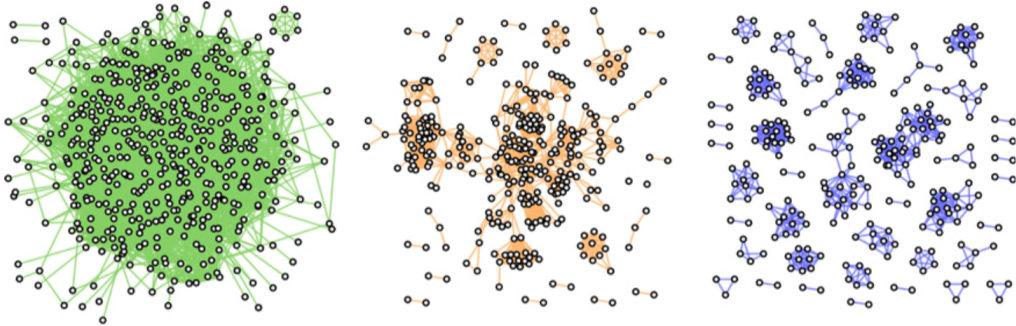}
\caption{Three views of the contact network. \textbf{Left} (green), all interactions aggregated over 24 hours. \textbf{Middle} (orange), interactions during one hour. \textbf{Right} (blue), interactions in a 5-minute window.  \label{fig:revealing_motifs}}
\end{figure}
The key realization arose from simply plotting the contact-network\cite[10mm]{stopczynski2014measuring} at increasingly higher temporal resolution.
The green hairball (Fig.\,\ref{fig:revealing_motifs}, left panel) shows connections between everyone who has spent time together, aggregated across an entire day. 
The orange network (Fig.\,\ref{fig:revealing_motifs}, middle panel) shows contacts aggregated over an hour, and the blue network (Fig.\,\ref{fig:revealing_motifs}, right panel) shows the interactions during a five-minute time slice. 
The discovery originates from the blue network. 
There, we can directly observe cross-sections of the fundamental structures: groups (well approximated by cliques) of people spending time together. 
We could directly observe the groups in the network without any need for community detection.

This was a case where analyzing the network became \emph{easier} by including higher resolution temporal data (in our case, no community detection was necessary).
Usually it is the opposite. 
Usually, our descriptions become more complex when we have to account for more detailed data, especially temporal data\cite{masuda2016guidance}.
I take the fact that \textit{more} data simplified this particular problem to mean that we were on to something: in this case, the fundamental structures constitute a quite natural representation of the network.
\begin{figure}
\centering
\includegraphics[width=0.9\textwidth]{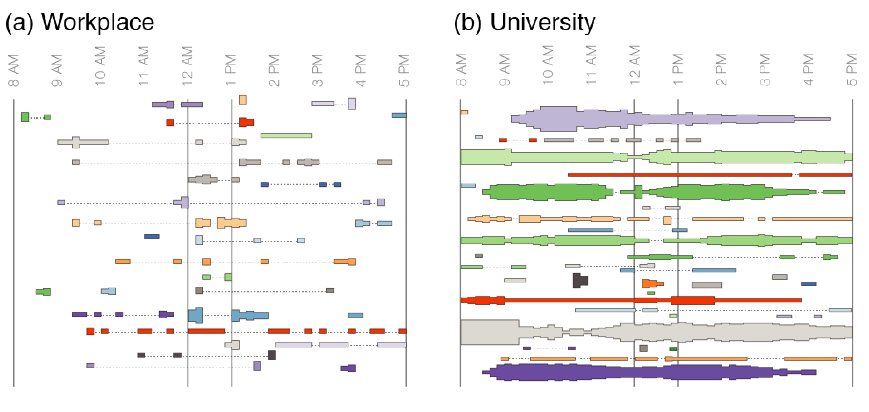}
\caption{An example of fundamental structures in \classtxt{real-time}, \classtxt{many}-\classtxt{to}-\classtxt{many} networks. In both panels, time runs from left to right, and each horizontal colored band represents a fundamental structure in that network (a sequence of cliques matched up over time). Therefore, each horizontal colored band is basically a representation of a group of people meeting, with the width of each band proportional to the number of participants at that time. Here we show these fundamental structures in two social settings (a) a Workplace network  (b) in the SensibleDTU data. \label{fig:sausages}}
\end{figure}
This way of representing the network also provided a way to understand the temporal evolution of the network.
Simply matching up cross-sections of the fundamental structures across time-slices, we could then construct the full fundamental structures (the individual communication events) for this class of network.
We called the result \emph{gatherings} -- the temporal representation of a meeting between a group of individuals (see Fig.\,\ref{fig:sausages} for an examples of gatherings in two real-world networks: a workplace\cite[23mm]{genois2015data} and university freshmen\cite{stopczynski2014measuring}).

Gatherings are the fundamental structure of \classtxt{many-to-many}, \classtxt{synchronous} networks.
Studying the properties of gatherings allowed us to estimate the relevant time-scales and spatial behaviors of the fundamental structures in this systems, e.g.~how individual nodes interact with the gatherings.
Turning our attention to time-scales of weeks and months, we could study the patterns of meetings (gatherings) among the same people beyond single meetings.
Thus, we could model the network dynamics as sequences of -- and relationships between -- such gatherings. 
This provided a dramatic simplification allowing us, for example, to make predictions about the temporal trajectories of individual nodes through the social network. 
We have since developed more sophisticated methods for identifying communities in this class of networks\cite{aslak2018constrained}.

\newthought{I include this example} to showcase the potential of the fundamental structures to organize our modeling of a certain network, and I hope that it will be possible to make similar progress for the remaining five dynamic classes.
Connecting to the more general point of within-class versus between-class comparisons, it is also important to emphasize, that while the descriptions and algorithms described above are excellent when analyzing networks in the \classtxt{synchronous}, \classtxt{many-to-many} class, they are not suited for describing networks in the remaining network classes (because they assume an underlying \classtxt{many-to-many}, \classtxt{synchronous} network structure).

\section{Frequently asked questions}
In this \textit{Frequently Asked Questions} (FAQ) section, I go over a few questions that have come up frequently when I have discussed the ideas in the paper with other researchers.

\subsection{What do you mean `framework'!?}
It is important to point out that the dynamic classes and associated fundamental structures are emphatically \textit{not} a mathematical framework (for example, the classes are neither disjoint, nor complete).
Instead my aim with this paper, is to point out new, meaningful structures in dynamic networks.
These structures are organized around the idea of \emph{communication events}, which in turn can be roughly classified into six prototypical forms of communication.
In this sense, aspects of the framework are qualitative, focusing on providing useful taxonomy of classes of networks in the real world. 

Nevertheless, as I argue in detail below, the fundamental structures impose a set of important constraints on dynamics for networks belonging to each class (with different constraints in different classes). 
These constraints impact many aspects of how we currently model and analyze temporal networks, and therein lies the value of the framework. 
Much more on this in the epilogue.

\subsection{Is the framework all done and ready to use?}
A very important point to make in this FAQ section is to admit that there is still a big piece of the framework missing. 
Specifically, that, while in the case of the \classtxt{synchronous} classes, understanding the temporal evolution of single communication events is relatively straightforward (as witnessed by our progress in the case of \classtxt{synchronous}, \classtxt{many}-\classtxt{to}-\classtxt{many} networks described above), the temporal structure of fundamental structures of networks from the \classtxt{asynchronous} classes is non-trivial since identification of (and method of analysis for) individual acts of communication is less clear. 

In these cases, for example, while there is a  well-defined end-time for a each fundamental structure (when the active node ceases to be available), structures themselves can still cease to show any link-activity much before that, for example an old Facebook post which it is technically possible to comment on, but which nobody will ever find again. Or a book, which nobody will ever read again, but which is still available on many bookshelves.
Further, in the \classtxt{many-to-many}, \classtxt{asynchronous} classes (which includes many important online social networks, such as Twitter and Facebook), there seems to be almost a spectrum running from \classtxt{one-to-many} to \classtxt{many-to-many}, depending on the amount of discussion associated with a post: posts without activity resembling trees, while vigorous discussions result in more clique-like structures.

\subsection{Is it just for communication networks?}
While we focus here on modeling communication networks, it is likely that the distinctions, concepts, and methods developed for each of the classes summarized in Table\,\ref{tab:matrix} are valid in domains outside human communication, for example dynamics of signaling networks in biology such as protein-protein interaction networks, gene regulatory networks, and metabolic networks. 
I also expect that the results developed in this project can be extended to networks of computer-to-computer communication.

\subsection{Isn't all this obvious?}
The distinctions pointed out in Table\,\ref{tab:matrix} may appear so self-evident that a reader might ask why they are currently not a part of modeling temporal networks.
I believe that the reason the network classes have remained unnoticed in the context of network science because time aggre\-ga\-tion has obscured the fundamental differences in \emph{generating processes} between networks with distinct fundamental structures.

\begin{figure*}
\centering
\includegraphics[width=0.9\textwidth]{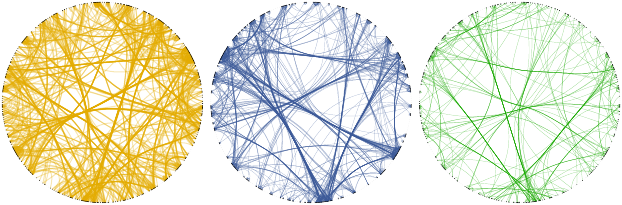}
\caption{Three networks defined on the same set of approximately 500 nodes from the SensibleDTU project, with links aggregated over one week; node-positions are the same in all three panels. From left to right the three panels show networks of physical proximity, telecommunication, and Facebook interactions. While from different dynamic classes, in aggregated form, all three networks have similar topological properties.\label{fig:aggregated}}
\end{figure*}

As noted above, at the level of aggregation used in the literature, the many distinct networks  (face-to-face, phone calls, text messages, emails, Twitter, Facebook, Snapchat, Instagram, discussion forums, etc.) that we participate in have common properties (see Fig.\,\ref{fig:aggregated}).
These common properties are due to the simple fact that all these networks reflect the same underlying object: the social network of relationships between human beings.
But as the cross sections of fundamental structures displayed in Figure~\ref{fig:revealing_motifs} shows, these networks are fundamentally different from each other on short time-scales.
These differences are due to the characteristics of (and design-choices behind) each communication platform, which inevitably encodes one of the prototypical forms of communication in Table\,\ref{tab:matrix}.

There are many traces of the fundamental structures in the recent literature. 
My group's work on communities in face-to-face networks discussed above proposes a new way of analyzing the class of \classtxt{synchronous}, \classtxt{many-to-many} networks, but does not realize its place in a larger framework. 
Elsewhere, recent work focusing on \textit{simplicial complexes} explores the same class, both in terms of network structures\cite[53mm]{petri2018simplicial} and implications for spreading processes\cite{iacopini2019simplicial}, again without noting that these networks are not necessarily representative of temporal networks generally; without explicitly pointing out that networks from different classes need different methods of analysis.
From another angle, it has been pointed out by many authors that time integrating techniques can introduce biases in understanding spreading processes\cite{krings2012effects, ribeiro2013quantifying} as we will discuss later.

\newthought{In the next section}, we explore the consequences of the presence of the six classes on selected topics within temporal network analysis.
Because each structure severely constrains possible network configurations, the fundamental motifs have a profound impact on the current state-of-the-art in temporal networks research.

\section{Consequences for analysis and modeling}
An immediate and important realization that flows from constraints imposed by the fundamental structures is that many important high-order network structures are strongly influenced by their network class.

I include an overview of five key topics below to illustrate the implications for existing temporal network theory.
This list is not exhaustive, but simply intended to give the reader some examples of where I think the dynamic classes could be useful for developing new descriptions of temporal networks.

\subsection{Randomization}
A common approach to understand the effect of temporal structure in networks is to use \emph{randomization} techniques to probe the impact of a structural feature of the network. 
A simple  example from static network theory to explain the logic of randomization:
In their seminal paper, Watts and Strogatz\cite{watts_wsmodel_1998} argued that real world networks are `small worlds', characterized by high clustering and short path lengths. 
But what does `high' and `short' mean in the sentence above? 
To make their point, Watts and Strogatz created `random' counterparts to their real-world networks which contained the same number of nodes and links as the empirical networks, but with links placed randomly among nodes.
They found that the empirical networks had both clustering and path-lengths that were orders of magnitude different (higher and lower, respectively) from their random counterparts.
In static networks, the degree distribution is also often conserved \cite{maslov2002specificity}.

The purpose of randomization is similar in temporal networks, but the possible randomization schemes are \textit{much} richer \cite{gauvin2018randomized}.
The idea is still: We want to estimate the effect of a specific temporal network property and remove that property (through randomization) to measure the effect.
One may shuffle time-stamps (to understand the importance of ordering), replace time-stamps with random times drawn from a uniform distribution (to understand the importance of circadian patterns), shuffle links (in order to destroy topological structures), reverse time (to understand importance of causal sequences), etc. 
The idea is then to simulate a process of interest on the temporal network and compare the dynamics of that process with the same process run on ensembles of networks that are increasingly randomized relative to the original network.

Because the fundamental structures (as I have argued above) correspond to individual communication events, it is not always meaningful to randomize the networks according to the strategies mentioned above -- \textit{this generally results in configurations of links that could not possibly appear in real communication networks}. Another, related issue is that the communication events (fundamental structures) themselves, often are the very thing that spread information/opinions. They are not always (as many modeling papers assume) an underlying infrastructure on which the spreading occurs. 

Thus, a fruitful area for future research is to develop  randomization schemes which respect the fundamental structures and understand how the fundamental structures impact the existing work on network randomizations.
A framework for randomization that respects the network classes would be analogous to the way that most randomizations in static networks respect degree distributions\cite{maslov2002specificity} (or higher-order structures\cite{orsini2015quantifying}), the key topological feature in these networks.

\subsection{Link prediction and link activity}
A dynamic network property strongly influenced by network class is the pattern of how links are active/non-active, and activity correlations between sets of links in a network\cite{karsai2012correlated}. 
In face-to-face networks, these patterns are typically dominated by long-duration meetings between groups of individuals (as discussed above), whereas in text message networks back-and-forth dynamics are common\cite{saramaki2015seconds}.

Closely related the link-activities is temporal link prediction \cite{liben2007link}. 
Here, the objective is to model patterns of link occurrences and use machine-learning to predict subsequent occurrences of links in the network based on local/global features of nodes/links. 
In static network theory, link prediction (especially within computer science) is a large topic\cite{lu2011link}, which focuses on predicting the presence of links that have been artificially removed or removed due to noise of some kind.
In temporal networks, the objective is often rephrased to -- for example -- predict all or some links in the next time-step\cite{dhote2013survey}. 

Based our understanding of the differences in link-activities in different classes, it is clear that the fundamental structures offer a way to understand why features for link-prediction can vary strongly from network to network. 
There is simply a massive difference between predicting future links in a \classtxt{synchronous}, \classtxt{many-to-many} network, where temporal cross-sections are cliques and structures typically persist for hours, relative to e.g.~text chat networks (\classtxt{asynchronous}, \classtxt{one-to-one}), where individuals can be in multiple ongoing conversations and text-snippets are short.
In turn, this means that link prediction algorithms trained on one class of networks will fare poorly on networks belonging to other classes, since features will change dramatically depending on network class.
These caveats become especially important when link-prediction is used to infer values for missing data\cite{clauset2008hierarchical, guimera2009missing}.

\newthought{Another consequence} for link prediction is that current performance estimations may be misleading.
This is because, depending on the dynamic class of network, not all links are possible to realize. 
\begin{figure}
\centering
    \includegraphics[width=0.9\textwidth]{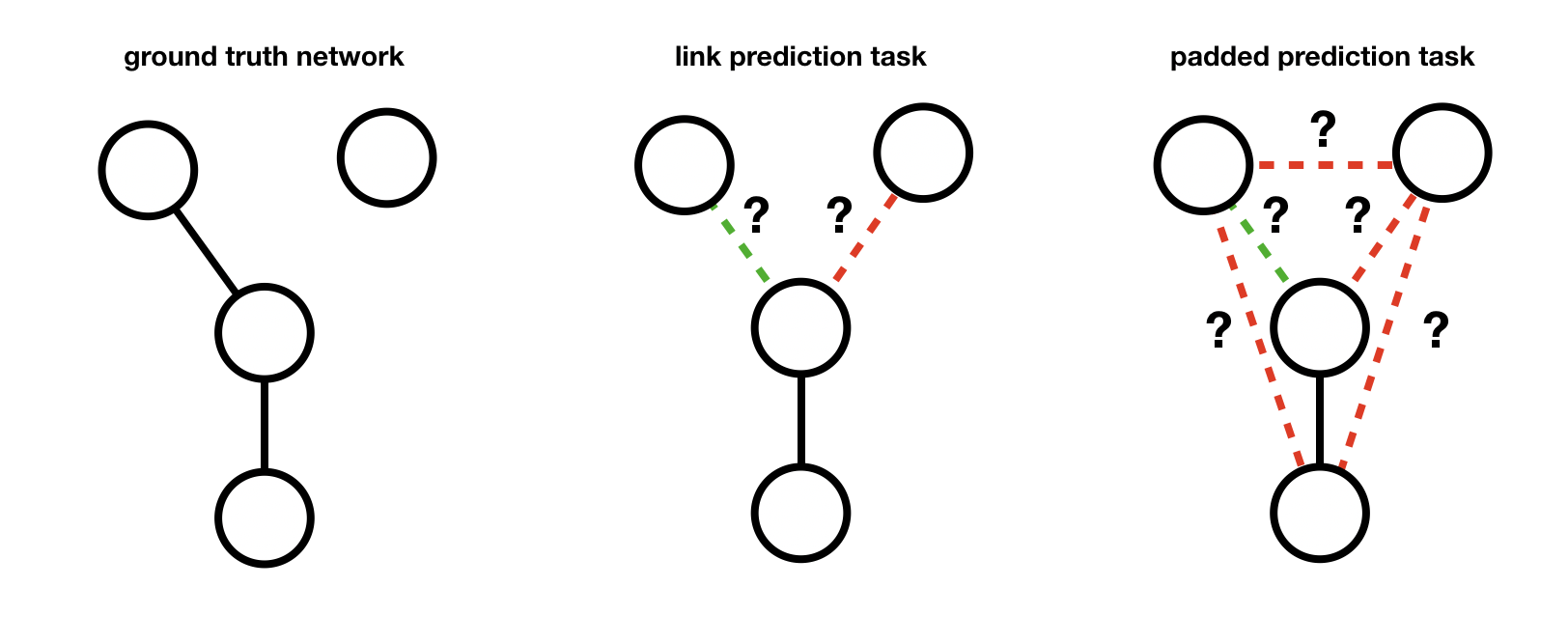}
\caption{Impossible links and link prediction performance. 
In this \classtxt{one-to-many} scenario, the only possible links connects central node to the three neighbors. 
In the left panel we see the ground truth network.
In the middle panel, we see the links that we are, in fact, relevant to consider when evaluating the performance of link-prediction.
In the rightmost panel, we show the `padded' network, which most current algorithms base their performance metrics on.
The padded task, however, includes a number of links that could not possibly occur. 
We are not interested in the classifier's performance on these links, and therefore an algorithm's ability to predict/not predict their presence, should not be a part of the performance evaluation.}
    \label{fig:link_prediction}
\end{figure}

When performing a link prediction task, we feed the classifier examples of removed links (`true' examples) and examples of links that never existed (`false' examples), we then evaluate whether the classifier can tell which links exist and which do not.
What we learn from the dynamic classes, is that there are, in fact, two types on non-links: actual false examples and `impossible' links -- links that cannot occur because they are not possible given the constraints imposed by fundamental structures in that network.
This problem is for example important in \classtxt{one-to-many} networks, where message recipients cannot communicate amongst each other, and there are many such impossible links.
Link prediction algorithms should only consider actual false examples and not the impossible links, see Figure\,\ref{fig:link_prediction} for an illustration of this problem in a \classtxt{one-to-many} network.

\subsection{Spreading processes}
Spreading processes are profoundly impacted by the fundamental structures. 
Let us begin the discussion on spreading by considering epidemic spreading.
Perhaps the most studied type of dynamical systems on temporal networks is epidemic spreading, realizing compartment models, such as SIS (susceptible-infected-susceptible), SIR (susceptible-infected-recovered), etc., on the temporal network.
In terms of disease spreading, the key quantity is the fraction of available Susceptible-Infected links at any given time. 
This fraction varies strongly depending on the network class\cite{mones2018optimizing}, which in turn means that we can expect spreading dynamics to unfold differently within different classes.

A central finding, for example, when simulating epidemics on temporal networks is that adding the temporal dimension has a strong impact on disease spreading in nearly all networks, relative to simulating the disease on a static network. 
In some cases the disease speeds up (relative to null models) and in others it slows down, depending on a complex interplay between structure and topology (see Holme's review on temporal networks\cite{holme2015modern} for a discussion).
This raises the intriguing possibility that perhaps some classes (e.g.~\classtxt{one-to-one} networks) might have slower epidemics than their randomized counterparts, while other classes (e.g.~\classtxt{many-to-many} networks) might have more rapidly spreading epidemics than their randomized versions.

\newthought{If we look beyond} epidemic spreading, there is experimental evidence that there are subtle differences in spreading processes across various domains and that opinions, behaviors, and information spread in different ways than diseases.
\begin{marginfigure}
\centering
    \includegraphics[width=\textwidth]{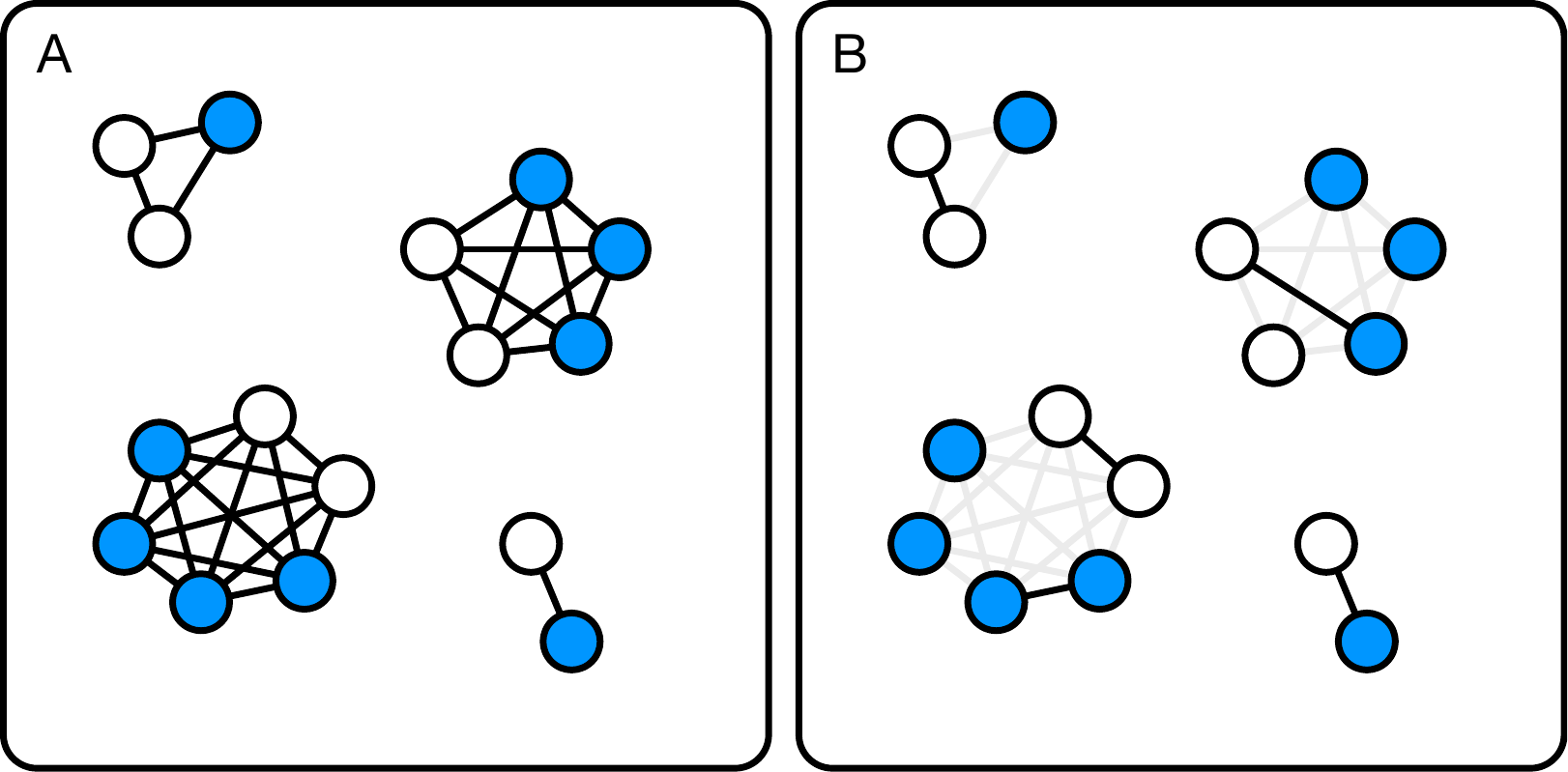}
\caption{A cartoon illustrating why complex contagion (e.g.~the threshold model) behaves differently in different classes of networks. A.~Shows the meetings in a many-to-many realtime network (face-to-face meetings). B.~Shows phone calls (\classtxt{one-to-one synchronous} network) among the same nodes at some point in time. Imagine that blue nodes are infected. In the many-to-many network, simultaneous information about a large set of neighbors is available for extended periods of time allowing for an accurate overview of opinions in the network. In the phone network nodes might need to wait extensively to access the state of some neighbors, allowing for much more difficulty in establishing an accurate state of knowledge.  \label{fig:threshold}}
\end{marginfigure}
When multiple sources of exposure to an innovation are required for a transmission to take place, we call the process \emph{complex contagion}.
The class of network has an even more profound impact on complex contagion processes than on simple disease spreading. 
Consider, for example, a threshold model\cite{granovetter1978threshold}, where the probability of infection depends on which fraction of a node's neighbors are infected. 
Compared to phone-call networks, for example, threshold models have fundamentally different outcomes in face-to-face networks, where large groups of individuals routinely gather\cite{iacopini2019simplicial}. 
In the phone call network, forming connections to a large fraction of one's network might take several months.
See Fig.~\ref{fig:threshold} for an illustration of this discussion.
Thus, if we want to understand contagion on a specific network, we must first understand the class of fundamental structures to which the network belongs.

\subsection{Communities}
Communities in static networks are groups of nodes with a high density of internal connections. 
Community detection in static networks never settled on a common definition of the term community (there is a strong analogy to clustering in machine learning\cite{kleinberg2003impossibility}).
Thus, generalizations to temporal networks also allow for substantial variability in approaches. 
The simplest strategy for identifying temporal communities is to first separate the list of time-stamped edges into sequence of static snapshots, independently cluster each layer, and then match the communities across the layers to find the temporal communities. 
A number of approaches can directly cluster the entire stack of temporal layers; these include three-way matrix factorization, time-node graphs , and stochastic block models.
See our recent paper\cite{aslak2018constrained} applying the InfoMap\cite{rosvall_infomap_2008} framework to find gatherings and cores in \classtxt{synchronous}, \classtxt{many-to-many} for detailed references on community detection in temporal networks.

From the perspective of the temporal structures, the central issue with community detection is that the appropriate community detection method varies strongly depending on a network's dynamic class. 
In \classtxt{synchronous}, \classtxt{many}-\classtxt{to}-\classtxt{many} networks, temporal continuity is a key feature of communities.
And as we have discussed above, communities in face-to-face networks (dynamic class: \classtxt{synchronous}, \classtxt{many-to-many}) form more or less instantaneously as a group of fully connected nodes that connect at a certain time and form gatherings that can be easily tracked over time.
In this sense, communities in face-to-face networks are straight-forward to identify -- they are literally the fundamental structures of such systems. 
Identifying communities in other dynamic classes, is a completely different exercise. 

Take Facebook (dynamic class: \classtxt{asynchronous} \classtxt{many}-\classtxt{to}-\classtxt{many}), as an example. 
In this class, communities become gradually observable as calls or messages aggregate over time.
In the latter case, communities have to do with other network properties than the temporal sequence.
In an asynchronous environment (such as an online social network), my interactions are driven by the order in which posts were published rather than organized by social context (as is the case in the \classtxt{synchronous} networks). 
To give a concrete example, I might retweet a work-related post about $p$-values, then `like' a post about the Finnish heavy metal band \emph{Nightwish}, published by a friend, and finally comment on a political statement from a family member.
Thus, in most \classtxt{asynchronous} systems, activity aggregates around \emph{active nodes} (posts) rather than social contexts. 
\emph{This means that interactions within communities are not necessarily correlated in time.}
A fact which must be taken into consideration when we construct methods for detecting communities.
At the same time, we know from the literature that communities do exist in these networks. 
So the question becomes, can we draw on fundamental structures to improve community detection in other classes?

As mentioned in the FAQ, the temporal evolution of the fundamental structures within the \classtxt{asynchronous} classes is under-determined in the framework as it currently stands. 
Similarly, exactly how to identify communities in these dynamical classes is not clear to me.
Therefore, the central point I wish to make related to communities, is simply that methods related to identifying communities in temporal networks will likely need to be different depending on the network's dynamical class.

\subsection{Generative Models}
Closely related to randomization is the idea of using the fundamental structures to build new synthetic networks.
The idea of using simple models that reproduce some properties of the system under study and its dynamics, has been another important method for understanding complex dynamical systems\cite{miller2009complex}. 
Realistic synthetic data is important because we can use such synthetic temporal networks to study dynamic processes. 
The synthetic networks provide access to arbitrary amounts of data where we (a)~understand the network's temporal changes (because we have created them) and (b)~create ensembles of networks to study variability in outcome given a particular dynamic (contrary to the case of real-networks, we typically only have a single instance).
Accordingly, a plethora of models that generate temporal networks have also been investigated.
See Holme's recent review\cite{holme2015modern} for an overview. 

Making now a sweeping (but I think true) statement, in the case of all these existing generative models, analyses based on synthetic datasets have little relevance for real-world problems \textit{because the models do not incorporate the constraints on dynamics imposed by the fundamental structures.}

Further, the framework of dynamic classes, however, offers a completely new way of generating synthetic temporal networks. 
This goes back to the idea of conceptualizing and modeling networks as sequences of communication events.
Since the fundamental structures are a manifestation of each network's real-world generative process, we can create new, synthetic network models by creating individual realistic communication events, then organizing those communication events a temporal sequence time to form the full network.
The usefulness of such models can be tested using statistical methods\cite{clegg2016likelihood}.

\section{Conclusion}
The lesson that i hope arises across the five examples above is that networks within each of the dynamic classes must be analyzed and modeled separately; that comparisons of statistics between networks are only meaningful for networks belonging to the same class. 
This is because the class itself (and not just the actual systems that are represented through the temporal network), strongly impacts almost all known temporal network metrics.

Zooming out further, three central lessons emerge from the full discussion of the dynamic classes and their fundamental structures. 
\begin{enumerate}
\item Firstly, I argued that it is meaningful to divide all communication networks into six dynamic classes (Fig.\,\ref{tab:matrix}).
This distinction originates from communication studies but is not yet recognized within network science.
\item Secondly, I pointed out that a network's class strongly influences its temporal evolution and alters dynamic processes on that network. 
This implies that we cannot meaningfully compare results for networks belonging to different classes.
\item Thirdly, I tried to motivate the idea that the dynamic classes provide a promising new framework for modeling temporal communication networks. 
This is because every communication network can be seen as \textit{sequences of individual communication events}. 
Thus, we can model every such network as generated by many instances of a single fundamental structure.
In this sense, the six classes provide us the foundation for a new framework for both measuring and modeling temporal networks.
\end{enumerate}
These three key take-homes lead me to consider the role that I hope the dynamical classes will play in the field of temporal networks.
An important element that is currently missing from the field of temporal network theory is a set of topological properties to measure and devise statistics for. 
This lack of agreed-upon-structures is eloquently pointed out by  Petter Holme in his excellent review of temporal network science\cite{holme2015modern}, where he writes:
\begin{quote}
\emph{In the history of static network theory, \textbf{measuring network structure has been driving the field}. For example, after Barab\'asi and coworkers discovered how common scale-free (i.e.~power-law-like) degree distributions are (\ldots), there was a huge effort both to measure degree distribution and to model their emergence.}

\emph{For temporal networks, \textbf{similar ubiquitous structures are yet to be discovered, perhaps they do not even exist}. This has led the research in temporal networks down a slightly different path, where the focus is more on dynamic systems on the network and how they are affected by structure, and less on discovering common patterns or classifying networks.} [my emphases]
\end{quote}
Now, allow me to speculate wildly for a bit. 
I do not think that it is impossible that the fundamental structures could be analogues to the `ubiquitous structures' mentioned in the quote for the case of temporal networks. 
Perhaps the six dynamic classes will allow us to think about structure in temporal networks in a new and more principled way. 

Finding such structures is important because, in static networks, a deeper understanding of the \emph{structure} of the network, has allowed us to reason in principled ways about their \emph{function} -- and for most applications outside pure science, function is what we care about. 
As the quote illustrates, temporal network science has had to follow a different path, focused more on simulation, for example observing how dynamical processes unfold. 
As a consequence, we still do not have a coherent picture of the key mechanisms in temporal networks. 
While still unproven at this point, I think that the fundamental structures carry the promise of being the ubiquitous structures that Holme posits are `yet to be discovered'. 
Therefore I hope that the new perspective provided by the dynamic classes will give rise to new statistical models, algorithms, and research questions.

\section*{Acknowledgements}
I would like to thank
Arkadiusz `Arek' Stopczynski,
Enys Mones, 
Hjalmar Bang Carlsen, 
Laura Alessandretti, 
James Bagrow,
Petter Holme,
Piotr Sapiezynski, 
Sebastiano Piccolo, 
Ulf Aslak Jensen, and 
Yong-Yeol Ahn
for fruitful discussions and generous comments on the manuscript text (list sorted alphabetically by first name). Special thanks to Piotr for the link prediction example.
This work was supported by the Independent Research Fund Denmark. The text was set using Tufte-\LaTeX\footnote{https://tufte-latex.github.io/tufte-latex/}.

\newpage
\section*{Epilogue: More FAQs}
There's a couple of more questions that have come up frequently in discussions of the framework, but which slowed down the flow of the paper, so I have moved them here, to the epilogue, for readers who might share these particular questions. 

\subsection*{What about mathematical completeness?}
A graph-theory inclined reader may to ask: `In what sense is this a mathematical framework?' With follow-ups such as `Are the classes disjoint? Can a dynamic network belong to multiple classes? Can a network's class change over time?' 
They might proceed `Are the classes complete? Can all possible networks be divided into one of the six classes? Is it possible to construct networks that fall outside the taxonomy in Table\,\ref{tab:matrix}?'
Here, the answer is: This is not a framework/theory in a graph theoretical sense. 
I think of the six classes as a \textit{model} in the physics sense of the word. 

Let me explain by way of an analogy. 
In the early days of quantum mechanics, Geiger and Marsden (directed by Rutherford), decided to shoot some $\alpha$-particles into a thin sheet of gold foil \cite{gegier1909diffuse, geiger1910scattering}.
They noticed that a vast majority of the particles went straight through the gold foil, but that a small fraction were scattered at a wide range of angles.
This was a highly unexpected and very non-classical behavior.
To explain these strange experimental observations, Rutherford proposed a new \emph{model}, qualitative at first, that atoms have a tiny and heavy nucleus, surrounded by a cloud of electrons (departing from the then popular `plum pudding model'\footnote{Yes, that was real thing.} of the atom, proposed in 1904 by J.J.~Thompson).
Based on Rutherford's model for the atom's structure, other scientists were able to develop better descriptions, eventually leading to the quantum mechanical framework that we teach undergrads today.

I think of the framework presented here as a model in the same sense as Rutherford's (no comparison otherwise).
Just like the model of a dense core with mostly empty space around it was a way to organize subsequent observations and provide structure to the theories/models to follow it, the dynamical classes are a way to organize our study of networks and to provide constraints/structure for the next steps of theory-building\footnote{By the way, as far as I can tell, the classes are not disjoint and not complete. Further, real networks are not necessarily a perfect fit to their classes. But as I hope to have convinced the reader by way of the analogy above \ldots that's not the point.}.

\subsection*{But how is this different from temporal motifs?}
Motifs are a structural characteristic closely connected to fundamental structures, and have been the focus of much research. 
This area features multiple generalizations of the motifs in static networks\cite{milo2002network} -- small subgraphs that occur more or less frequently than one might expect in an appropriate null model.
Typically, the strategy is to count the temporal subgraphs occurring within some interval $\Delta t$.\cite{kovanen2013temporal}
Findings suggest that certain tit-for-tat motifs and triangles are over-represented in phone networks (dynamic class: \classtxt{synchronous}, \classtxt{one-to-one} networks) and may shape processes such as spreading.
Other motif-like structures have been explored, for example graphlets, which are equivalence classes of $\Delta t$-causal subgraphs.
Of particular relevance to the framework presented here is work on structure prediction and related algorithms for efficiently counting isomorphic temporal subgraphs.
For a more detailed discussion and references, I once again refer the reader to recent reviews\cite{holme2012temporal, holme2015modern} .

From the perspective of fundamental structures, there are two issues with temporal sub-graph counting approaches.
The key issue is that current methods do not measure individual communication events.
The sliding window based approach, which identifies the network structures that arise within some time $\Delta t$ does not recognize that the fundamental structures have a natural beginning and end.
As a consequence, these methods do not identify and aggregate statistics for the fundamental structures, rather ending up with aggregate statistics for smaller structures which are incidental to the fundamental structures. 

\newpage

\bibliography{bibliography}
\bibliographystyle{plainnat}

\end{document}